# Liquidity Jump, Liquidity Diffusion, and Treatment on Wash Trading of Crypto Assets

Qi Deng[1,2,3,4*]
Zhong-guo Zhou[5]

## Abstract

We propose that the liquidity of an asset includes two components: liquidity jump and liquidity diffusion. We show that liquidity diffusion has a higher correlation with crypto wash trading than liquidity jump and demonstrate that treatment on wash trading significantly reduces the level of liquidity diffusion, but only marginally reduces that of liquidity jump. We confirm that the autoregressive models are highly effective in modeling the liquidity-adjusted return with and without the treatment on wash trading. We argue that treatment on wash trading is unnecessary in modeling established crypto assets that trade in unregulated but mainstream exchanges.

Funding Source: The work was supported by Hubei University of Automotive Technology under Grant number BK202209]; Hubei Provincial Bureau of Science and Technology under Grant number 2023EHA018].

1. College of Artificial Intelligence, Hubei University of Automotive Technology, Shiyan, China
2. Jack Welch College of Business & Technology, Sacred Heart University, Fairfield, CT, USA
3. School of Accounting, Economics and Finance, University of Portsmouth, Portsmouth, UK
4. Cofintelligence Financial Technology Ltd., Hong Kong and Shanghai, China
5. Department of Finance, Financial Planning, and Insurance, Nazarian College of Business and Economics, California State University Northridge, CA, USA
*. Corresponding author: dq@huat.edu.cn; dengq@sacredheart.edu

# Liquidity Jump, Liquidity Diffusion, and Treatment on Wash Trading of Crypto Assets

## 1. Introduction

Liquidity risk is a crucial factor in asset trading. When an asset lacks liquidity, it can lead to wider bid-ask spreads, increased transaction costs, and potential price volatility. On the other hand, extreme liquidity leads to price jumps, causing the asset's return and volatility to lose their autoregressive properties and therefore become hard to model by the mainstream autoregressive models. For example, Deng and Zhou (2024) demonstrate that the traditional ARMA-GARCH/EGARCH models are no longer effective in modeling assets with extreme liquidity, but that if the return and volatility of the assets are adjusted by proper liquidity measures, their well-behaved autoregressive properties can be restored, and they can be modeled effectively by these models again. They use crypto assets to represent the assets with extreme liquidity, of which the trading is unregulated and therefore subject to a manipulative trading practice known as "wash trading" (Cong et al., 2023).

In order to combat the effect of wash trading, in their attempt to model the liquidity of crypto assets, Deng and Zhou (2024) deploy treatment. In the process of aggregating the tick-level data (collected from Binance) to calculate the minute-level (intraday) trading amount ($A_t$), they divide $A_t$ into four quantiles (Q1-Q4) of equal quantity and reduce the quantity of Q3 (50 percentile) by a factor of 50%, and the quantity of Q4 (75 percentile) by a factor of 75%. From the minute-level data they further calculate daily amount, which is on average about 40% of the untreated "raw" daily amount across the selected assets, or about 60% of the raw daily amount is regarded as being from wash trades. Since Cong et al. (2023) estimate that wash trades count for about 46.47% of the total amount in Binance, the treatment of Deng and Zhou (2024) is actually more stringent.

This paper is inspired by Deng and Zhou (2024) and Cong et al. (2023). Cong et al. (2023) find that the wash trading of a crypto asset is proportional to the asset's price over the immediate horizon (during the same week), which is consistent with the intention of wash trading initiators to induce "short-term price jumps" and benefit from them. This suggests that treatment on wash trading that reduces price jumps should improve the asset's responsiveness to the autoregressive models. However, Deng and Zhou (2024) find that the liquidity-adjusted ARMA-GARCH/EGARCH constructs are equally effective in modeling crypto assets, either with or without such treatment. The apparent contradiction prompts us to raise an important question: is it necessary to provide treatment on wash trading in modeling the return and volatility of crypto assets? By answering this question, we establish a proper framework for our subsequent studies, and more importantly, we expand the current literature in providing an alternative methodological approach to modeling crypto assets, and in general, any assets with extreme liquidity.

In order to answer the above question and based on the findings of Deng and Zhou (2024) that the autoregressive properties can be restored to the return and volatility of assets with extreme liquidity if they are adjusted with proper liquidity measures. In particular, we need to resolve the following two key issues: 1) what liquidity measure is actually affected by the treatment on wash trading, and 2) whether the effect of treatment on the liquidity measure is adequate to restore the responsiveness of the asset return and volatility to the autoregressive models. As such, and based on our observations, we propose the following two hypotheses: 1) that the treatment on wash trading in Deng and Zhou (2024) only reduces the volatility of jumps, but not the magnitude of jumps that actually dominate the asset's responsiveness to the autoregressive models; and 2) that the liquidity adjustment in Deng and Zhou (2024) alone is adequate in reducing the magnitude of price jumps with and without the treatment on wash trading, and therefore is equally effective under either condition in restoring the asset's responsiveness to the autoregressive models.

The above hypotheses are the direct reasons for us to investigate whether liquidity itself can be decomposed into two distinctive yet complementary components that react to the treatment on wash trading differently: a "liquidity jump" that represents the magnitude of jumps and is believed insensitive to the treatment, and a "liquidity diffusion" that reflects the volatility of jumps and can be greatly affected by the treatment. The liquidity jump is defined as the ratio of regular return and liquidity-adjusted return that measures the magnitude of daily liquidity in a given day, while the liquidity diffusion is defined as the ratio of regular volatility and liquidity-adjusted volatility that reflects the intraday volatility of the daily liquidity. The definitions of liquidity-adjusted return and volatility are given by Deng and Zhou (2024). We provide empirical evidence to support Hypothesis 1 by demonstrating that treatment on wash trading significantly reduces the level of liquidity diffusion, but only marginally reduces the level of liquidity jump to a degree that is inadequate to restore the asset's responsiveness to the autoregressive models.

Following Deng and Zhou (2024), we apply the forecasted daily liquidity-adjusted return produced by the liquidity-adjusted ARMA-GARCH/EGARCH models as the inputs to the LAMV constructs for portfolio optimization.[1] For comparison, we duplicate the procedure with the forecasted daily regular (non-liquidity-adjusted) return being fed into the TMV constructs. We observe a clear advantage for the LAMV over the TMV in portfolio performance, either with or without treatment on wash trading. Our empirical evidence supports Hypothesis 2 that liquidity adjustment significantly reduces the level of liquidity jump regardless of treatment on wash trading. Combining the empirical evidence that support both hypotheses, we conclude that liquidity adjustment to the return and volatility of assets with extreme liquidity reduces the level of liquidity jump adequately in restoring the autoregressive properties, while that treatment on wash trading is not necessary as it does not adequately reduce the level of liquidity jump.

---

[1] Following Deng and Zhou (2024), we use the standard Mean-Variance (MV) method to optimize portfolios with forecasted returns produced by the liquidity-adjusted ARMA-GARCH/EGARCH models and name these portfolios the "liquidity-adjusted Mean Variance (LAMV)" portfolios. We also use the standard MV to optimize portfolios with forecasted returns produced by the traditional ARMA-GARCH/EGARCH models and name these portfolios the "traditional Mean Variance (TMV)" portfolios. We use these terms throughout this paper.

The contribution of the paper is four-fold. First, we propose two distinctive yet complementary asset-level liquidity measures: the liquidity jump that measures the magnitude of aggregated price jumps in a given day, and the liquidity diffusion that reflects the intraday volatility of the aggregated daily jumps. Second, we establish that treatment on wash trading greatly reduces the level of liquidity diffusion but does not adequately reduce the level of liquidity jump, which actually dominates the asset's responsiveness to the autoregressive models. Third, we demonstrate that liquidity adjustment to the return and volatility of assets with extreme liquidity adequately reduces the level of liquidity jump with and without treatment on wash trading, and therefore is effective in restoring the asset's responsiveness to the autoregressive models under either condition. Therefore, treatment on washing trading for established crypto assets is neither adequate nor necessary. And forth, we provide a simple and practical investment strategy that established crypto assets traded in mainstream exchanges may be considered as a viable alternative asset class for portfolio managers, if their return and volatility are adjusted by proper liquidity measures. While we use crypto assets to exemplify the usefulness and utilities of our models, they are effective on other assets with extreme liquidity.

The rest of the paper proceeds as follows. Section 2 reviews literature on components of asset liquidity. Section 3 introduces the asset-level liquidity jump and liquidity diffusion based on the liquidity-adjusted return and volatility proposed by Deng and Zhou (2024). Section 4 provides descriptive statistics and visualizations of the dataset and discusses the distributions of liquidity jump and diffusion. Section 5 reviews the liquidity-adjusted ARMA-GARCH/EGARCH models in Deng and Zhou (2024). Section 6 optimizes the portfolios of selected crypto assets with the MV specifications to provide empirical support to our hypotheses. Section 7 concludes the paper.

## 2. Literature Review

Deng and Zhou (2024) provide a thorough literature review on asset-level liquidity measures, liquidity costs, volatility of liquidity, and models of assets with extreme liquidity. In this section, we briefly review the literature (or lack of) on modeling asset liquidity in the context of (explicitly) dividing liquidity into two distinctive components: liquidity jump and liquidity diffusion and discuss the impact of each component on asset trading. In addition, we review the literature that addresses wash trading of unregulated assets with extreme liquidity, i.e., crypto assets.

Through a careful search and to our best knowledge, this paper is the first attempt to explicitly decompose liquidity into liquidity jump and liquidity diffusion. Some earlier work, such as those by Andersen et al. (2001), Amihud (2002), and Hou and Moskowitz (2008) show liquidity can vary across different market conditions and asset classes, which, in essence, discuss the cross-sectional variability of liquidity jump. Gabaix et al. (2006) find evidence of clustering in jumps, that extreme liquidity is not purely random but driven by common factors or dynamics in the market, suggesting the temporal variability of liquidity jump. Another stream of research, such as Datar, Naik, and Radcliffe (1998), Chordia and Subrahmanyam (2004), and Bekaert, Harvey and Lundblad (2007) show that the level of liquidity variability is generally associated with transaction costs, bid-ask spreads, and market depth, touching upon the liquidity diffusion in the context of market microstructure. Roll (1984), Huberman and Halka (2001), and Aït-Sahalia, Mykland and Zhang (2005) share a broader view on a crucial role that liquidity plays towards market efficiency, pricing dynamics, and risk management in continuous-time processes, revealing both the temporal and cross-sectional nature of liquidity jump and liquidity diffusion.

However, there are two major gaps in the abovementioned literature: 1) there is no explicit attempt to further divide liquidity into a jump component and a diffusion component, and therefore the impact of liquidity on the trading of assets with extreme liquidity cannot be fine-tuned; and 2) all studies have an (implicit) assumption that the assets under investigation are strictly-regulated and well-behaved that do not exhibit frequent extreme liquidity and are not subject to wash trading. Both gaps render their models inadequate in dealing with the trading of crypto assets.

Le Pennec, Fiedler and Ante (2021) analyze wash trading based on web traffic and wallet data and propose that wash trading counts for more than 90% volume-wise for most investigated crypto exchanges. Aloosh and Li (2024) provide evidence for the wash trading of Bitcoin across different crypto exchanges. Sifat, Tariq and van Donselaar (2024) examine wash trading in nonfungible token (NFT) transactions. The most comprehensive study on crypto wash trading is Cong et al. (2023), which categorizes crypto exchanges into two groups: a regulated one and an unregulated one (two tiers) and demonstrates that wash trades average more than 70% of the reported volumes for the unregulated exchanges. These studies establish that the potential impact of wash trading, in the context of asset liquidity, must be adequately analyzed and/or addressed. However, while these important studies deploy sophisticated methodologies in documenting and statistically estimating the size of wash trades among crypto assets, they do not offer countermeasures or other tools to actually mitigate the effect of wash trading in modeling crypto assets, and therefore do not convey clear messages to investors on what they should actually do to combat wash trading.

## 3. Liquidity Jump and Liquidity Diffusion

In this subsection, we first briefly review the asset-level liquidity-adjusted return and volatility proposed by Deng and Zhou (2024). Using the tick-level trading data from the most dominant crypto asset exchange, Binance, Deng and Zhou (2024) model the unobservable minute-level liquidity-adjusted volatility $\sigma^2{}_T^{\ell}$ and return $r_t^{\ell}$ at equilibrium for time-period $T$ (a 24-hour/1440-minute trading day) as follows:

$$\sigma^2{}_T^{\ell} = \frac{1}{T}\sum_{t=1}^{T} \eta_T \frac{|r_t|/\overline{|r_t|}}{A_t/\overline{A_t}} (r_t - \bar{r}_t)^2 = \frac{1}{T}\sum_{t=1}^{T}\left(r_t^{\ell} - \overline{r_t^{\ell}}\right)^2$$

$$r_t^{\ell} = \sqrt{\eta_T \frac{|r_t|/\overline{|r_t|}}{A_t/\overline{A_t}}}\, r_t$$

$$subject\ to:\ \sum_{t=1}^{T} \eta_T \frac{|r_t|/\overline{|r_t|}}{A_t/\overline{A_t}} = T \Rightarrow \eta_T = \frac{T}{\sum_{t=1}^{T} \frac{|r_t|/\overline{|r_t|}}{A_t/\overline{A_t}}}$$

$$\Rightarrow \beta_t^{\ell} = \frac{r_t}{r_t^{\ell}} = \frac{1}{\sqrt{\eta_T \frac{|r_t|/\overline{|r_t|}}{A_t/\overline{A_t}}}}$$

*where*:
1. $r_t$ is the observed return in minute $t$, $\bar{r}_t$ is its arithmetic average in that day,
2. $|r_t|$ is the absolute value of $r_t$, $\overline{|r_t|}$ is its arithmetic average in that day,
3. $r_t^{\ell}$ is the liquidity-adjusted return in minute $t$, $\overline{r_t^{\ell}}$ is its arithmetic average in that day,
4. $A_t$ is the dollar (USDT) amount traded in minute $t$, $\overline{A_t}$ is its arithmetic average in that day,
5. $\eta_T$ is the daily normalization factor on day $T$ and is a constant for day $T$,
6. $\beta_t^{l}$ is the liquidity premium factor in time $t$,
7. $T = 1440$ as there are 1440 minutes (24 hours) in a crypto asset trading day.

The minute-level "liquidity premium Beta" $\beta_t^{\ell}$ is a unitless and normalized liquidity measure:

$$\beta_t^{\ell} \subset \begin{cases} > 1; high\ liquidity \\ = 1; equilibrium\ liquidity \\ < 1; low\ liquidity \end{cases}$$

The daily (day-level) regular and liquidity-adjusted returns for time-period $T$ is obtained by aggregating the intraday (minute-level) returns are given as:

$$r_{TT} = (1 + r_t)^T - 1 \cong (1st\ order)\ \sum_{t=1}^{T} r_t$$
$$r_{TT}^{\ell} = \left(1 + r_t^{\ell}\right)^T - 1 \cong (1st\ order)\ \sum_{t=1}^{T} r_t^{\ell}$$

The realized and unobservable daily (intraday on minute-level) variance for time-period $T$ is:

$$\sigma^2{}_{TT}^{\ell} = T\sigma^2{}_T^{\ell}$$

The "daily liquidity premium Beta," $\beta^{\ell}_{r_{TT}}$, for time-period T is thus defined as follows:

$$\beta^{\ell}_{r_{TT}} = |r_{TT}/r^{\ell}_{TT}| \subset \begin{cases} > 1; high\ daily\ liquidity\ magnitude \\ = 1; equilibrium\ daily\ liquidity\ magnitude \\ < 1; low\ daily\ liquidity\ magnitude \end{cases} \quad (1)$$

In this paper, we expand Deng and Zhou (2024) and introduce a "daily liquidity volatility Beta," $\beta^{\ell}_{\sigma_{TT}}$, which reflects the volatility of liquidity for time-period T, defined as follows:

$$\beta^{\ell}_{\sigma_{TT}} = |\sigma_{TT}/\sigma^{\ell}_{TT}| \subset \begin{cases} > 1; high\ daily\ liquidity\ volatility \\ = 1; equilibrium\ daily\ liquidity\ volatility \\ < 1; low\ daily\ liquidity\ volatility \end{cases} \quad (2)$$

Together, $\beta^{\ell}_{r_{TT}}$ and $\beta^{\ell}_{\sigma_{TT}}$ form the "daily liquidity Beta" pair, which enables us to further divide the asset-level liquidity into two distinctive yet complementary components: liquidity jump and liquidity diffusion. We regard $\beta^{\ell}_{r_{TT}}$ as the "liquidity jump," which serves as a proxy of liquidity magnitude that indicates extreme liquidity when its value is much greater than ($\beta^{\ell}_{r_{TT}} \gg 1$). At the same time, we regard $\beta^{\ell}_{\sigma_{TT}}$ as the "liquidity diffusion," which reflects liquidity volatility that points to extreme liquidity volatility when its value is much greater than 1 ($\beta^{\ell}_{\sigma_{TT}} \gg 1$).[2]

## 4. Dataset, Descriptive Statistics, and Distribution Visualizations

We follow Deng and Zhou (2024) to select 10 largest crypto assets by their market values with at least 4.5 years of historical data (April 27, 2019 to February 8, 2024 with 1,749 trading days) for 3.5-year back-tests (April 27, 2020 to February 8, 2024, 1,383 trading days) with a 1-year (365 days) rolling window. The selected crypto assets are ADA, BNB, BTC, ETC, ETH, LINK, LTC, MATIC, XMR and XRP. We first collect tick-level trading data of these assets from the Binance API and aggregate the tick-level data to construct minute-level (intraday) amount. We then construct minute-level (intraday) return and variance, both regular ($r_t$ and $\sigma^2_t$) and liquidity-adjusted ($r_t^\ell$ and $\sigma^2{}_t^\ell$). From the minute-level data we calculate their corresponding daily returns and variances, both regular ($r_{TT}$ and $\sigma^2{}_{TT}$) and liquidity-adjusted ($r_{TT}^\ell$ and $\sigma^2{}_{TT}^\ell$), as well as the daily liquidity premium Beta $\beta_{r_{TT}}^\ell$ and daily liquidity volatility Beta $\beta_{\sigma_{TT}}^\ell$. We conduct the same procedure twice to create two sets of data, one with treatment on wash trading and one without, in order to study the impact of treatment on wash trading.[3] We report the descriptive statistics of $\beta_{r_{TT}}^\ell$ and $\beta_{\sigma_{TT}}^\ell$ for all ten assets in Tables 1 to 2, with companion visualizations in Figures 1 to 6.

---

[2] Practically, liquidity magnitude is "extreme" when $\beta^{\ell}_{r_{TT}} \geq 4$, and liquidity volatility is "extreme" when $\beta^{\ell}_{\sigma_{TT}} \geq 2.5$.

## 4.1 Comparisons of $\beta_{r_{TT}}^\ell$ with and without Treatment on Wash Trading

Panel A of Table 1 summarizes the descriptive statistics of the daily liquidity premium Beta $\beta_{r_{TT}}^\ell$ for all the assets with treatment on wash trading, and Panel B of Table 1 provides similar statistics without treatment on wash trading. Comparing the descriptive statistics from Panel A with those from Panel B, we observe that, the mean (between 1.19 for BTC and 1.57 for ADA) and median (ranging from 0.73 for BTC to 0.91 for BNB) of $\beta_{r_{TT}}^\ell$ with treatment are all reduced from mean (between 1.40 for BTC and 2.02 for ADA) and median (ranging from 0.90 for XMR to 1.06 for ADA) without treatment, respectively. On the other hand, although the number of days with the maximum value of $\beta_{r_{TT}}^\ell$ (=10) is reduced to a range of 30 for BTC to 64 for ADA with treatment from a range of 44 for BTC to 110 for XMR without treatment, its share in the overall $\beta_{r_{TT}}^\ell$ load is reduced by a much smaller margin, to a range of 14.40% for BTC to 23.81% for XMR with treatment from a range of 16.85% for BTC to 31.81% for XMR without it. It seems that treatment on wash trading removes some extreme liquidity and thus reduces the level of liquidity jump by a certain degree, but it is not adequate to alter the asset's reaction to modeling.

---

[3] We follow the method of removing wash trades by Deng and Zhou (2024).

## 4.2 Comparisons of $\beta^{\ell}_{\sigma_{TT}}$ with and without Treatment on Wash Trading

Panels A and B of Table 2 summarize the descriptive statistics of the daily liquidity volatility Beta $\beta^{\ell}_{\sigma_{TT}}$ for all the assets with and without treatment on wash trading. The mean (between 0.66 for BTC and 0.87 for ETC and XMR) and median (ranging from 0.67 for BTC to 0.88 for XMR) of $\beta^{\ell}_{\sigma_{TT}}$ with treatment are all reduced from the mean (between 0.87 for ETH and 1.49 for XMR) and median (ranging from 0.82 for BTC to 1.18 for ETC) without treatment. It is interesting that the number of days with the maximum value of $\beta^{\ell}_{\sigma_{TT}}$ (=10) is reduced to 0 for all the assets with the treatment from a range of 0 for BNB, BTC, ETC, ETH to 19 days for XMR without the treatment. What is more significant is that with the treatment on wash trading, there is no day with extreme liquidity volatility ($\beta^{\ell}_{\sigma_{TT}} \gg 1$) as the range of maximum $\beta^{\ell}_{\sigma_{TT}}$ is between 0.85 for BTC and 2.07 for LINK, while without the treatment, only BTC ($max = 1.15$) and ETH ($max = 1.45$) have no day with extreme liquidity volatility. It seems that the treatment on wash trading completely removes trades that cause extreme liquidity volatility (which admittingly are not many to begin with), and therefore greatly reduces the level of liquidity diffusion.

## 4.3 Visualizations of $\beta^{\ell}_{r_{TT}}$ and $\beta^{\ell}_{\sigma_{TT}}$ with and without Treatment on Wash Trading

Figures 1 and 2 provide the scatter plots of $\beta^{\ell}_{r_{TT}}$-$\beta^{\ell}_{\sigma_{TT}}$ for all the assets with and without the treatment on wash trading respectively. The most obvious observation is that a plot in Figure 1 (with the treatment) has a much narrower interval along the x-axis ($\beta^{\ell}_{\sigma_{TT}}$) than its corresponding plot in Figure 2 (without the treatment) with the noticeable exceptions of BTC and ETH, while both plots have the same interval along the y-axis ($\beta^{\ell}_{r_{TT}}$) with a maximum value of 10. These visualizations illustrate that the treatment on wash trading indeed removes a number of jumps and all extreme liquidity volatility, resulting in a marginally lower level of liquidity jump and a significantly lower level of liquidity diffusion.

Figures 3 and 4 provide additional visualizations for liquidity jump $\beta^\ell_{r_{TT}}$ with and without the treatment on wash trading respectively. In Figure 3 with the treatment, a set of histograms (Figure 3 Column 1) confirms that $\beta^\ell_{r_{TT}}$ has a thin yet very long right tail for all the assets.[4] Also a set of scatter plots ($r_{TT}$ vs. $r^\ell_{TT}$) overlapping a straight line with a coefficient of 1.0 ($\beta^\ell_{r_{TT}} = 1.0$ or equilibrium value) (Figure 1 Column 2) confirms that there are more days with low liquidity ($\beta^\ell_{r_{TT}}$ < 1.0) than the days with high liquidity ($\beta^\ell_{r_{TT}}$ > 1.0). Furthermore, a set of 3D scatter plots in Figure 1 Column 3 illustrates the distribution of $r_{TT}$-$r^\ell_{TT}$-$\beta^\ell_{r_{TT}}$, which provides additional information from the third dimension of liquidity ($\beta^\ell_{r_{TT}}$). In summary, the liquidity jump $\beta^\ell_{r_{TT}}$ is highly asymmetric, positively skewed with a long right tail for all the assets. The results are consistent with Deng and Zhou (2024). In Figure 4 (without treatment), the histograms (Figure 4 Column 1) show that the distribution of $\beta^\ell_{r_{TT}}$ is similar to that of its counterpart in Figure 3 (with treatment) for all the assets, but with a "taller" bar on the far right ($\beta^\ell_{r_{TT}} = 10$), indicating that the treatment removes a number of trades with extreme liquidity that are presumed to be wash trades.

Figures 5 and 6 provide additional visualizations for liquidity diffusion $\beta^\ell_{\sigma_{TT}}$, with and without the treatment on wash trading. With the treatment (Figure 5), $\beta^\ell_{\sigma_{TT}}$ has a relatively symmetric and narrow distribution with no extreme values (Figure 5 Column 1). Also a set of scatter plots ($\sigma_{TT}$ vs. $\sigma^\ell_{TT}$) overlapping a straight line with a coefficient of 1.0 ($\beta^\ell_{\sigma_{TT}}$=1.0 or equilibrium value) (Figure 5 Column 2) confirms that an overwhelming majority of days has low liquidity volatility ($\beta^\ell_{\sigma_{TT}}$ <

---

[4] There are spikes of $\beta^\ell_{r_{TT}}$ at 10, as again we cap them to avoid extremely large values.

1.0). Furthermore, the 3D scatter plots in Figure 5 Column 3 provide distributions of $\sigma_{TT}$-$\sigma_{TT}^{\ell}$-$\beta_{\sigma_{TT}}^{\ell}$ to visualize the additional information from the third dimension of liquidity ($\beta_{\sigma_{TT}}^{\ell}$). The histograms in Figure 6 Column 1 (without treatment) show that the distribution of $\beta_{\sigma_{TT}}^{\ell}$ is different from that of its counterpart in Figure 5 (with treatment), evidenced by a significantly longer right tail and a bar on the far right ($\beta_{\sigma_{TT}}^{\ell} = 10$), indicating that the treatment removes almost all the trades with extreme liquidity volatility potentially brought by wash trades, with the noticeable exceptions of BTC and ETH.

### 4.4 Discussions on $\beta_{r_{TT}}^{\ell}$ and $\beta_{\sigma_{TT}}^{\ell}$ as Indicators of Wash Trading

It is worth mentioning that the two most popular crypto assets, by either market value or media popularity, BTC (Bitcoin) and ETH (Ethereum), do not have extreme liquidity volatility even without the treatment of wash trading (1.15 and 1.45, respectively). This observation, combined with the fact that BTC and ETH also have the lowest number of days with the maximum value of $\beta_{r_{TT}}^{\ell}$ (44 days and 52 days from Panel B of Table 1, respectively) without the treatment, indicates that BTC and ETH have fewer number of trades with relatively high volumes compared to other crypto assets that are unlikely to be wash trades.[5]

From a technical standpoint, the above observations provide an alternative method of measuring the severity of wash trading, that the liquidity diffusion $\beta_{\sigma_{TT}}^{\ell}$ has a higher correlation with wash trading than the liquidity jump $\beta_{r_{TT}}^{\ell}$. A reasonable economic explanation is that, when trading large market-cap crypto assets in established exchanges such as Binance, manipulative traders do not conduct a small number of very large-volume trades, but engage in high-frequency, large number of small-volume and momentum trades, sometimes with drastically different price points, as their motivation is not necessarily the long-term upward price movement that they may have difficulty to maintain, but small and frequent short-term gains that are sustainable. These trades in turn increase liquidity volatility, resulting in higher levels of liquidity diffusion. On the other hand, the vast majority of very large-volume trades that induce higher level of liquidity jump may actually reflect legitimate high demand compounded with unregulated trading. Our results indicate that treatment on wash trading essentially removes all small-volume wash trades, but only some large-volume wash trades.

---

[5] Note that in the context of crypto trading a "high volume" is relative, that a relatively small volume for BTC may still be larger in absolute term than a relatively large volume for a crypto asset with a smaller market cap, however, the former may have a lower liquidity premium Beta $\beta_{r_{TT}}^{\ell}$ than the latter.

The above results seem to suggest that liquidity diffusion is a better indicator of wash trading than the liquidity jump, and a combination of high liquidity jump, and high liquidity diffusion is the most reliable indicator for wash trading. A closer look of the scatter plots in Figure 2 (without treatment) reveals that that the number days with both high liquidity jump and high liquidity diffusion, presumably resulting from wash trades, is very scarce (between 0 and ~10) for all assets. Therefore, we argue that wash trading is actually very uncommon among high market-cap crypto assets (even beyond BTC and ETH) that trade in established exchanges, and certainly not to the degree suggested by Cong et al. (2023). The reason may reside in that Cong et al. (2023) investigate a large number of smaller exchanges that trade a substantial number of low-market-cap crypto assets, and smaller exchanges that trade practically worthless crypto assets ("air coins") are more prone to wash trading. In our study, for the selected 10 crypto assets with high market-cap, long trading history, and traded in the top exchange, Binance, wash trading is insignificant. In addition, treatment on wash trading might actually reduce portfolio performance if legitimate high-volume trades are removed (see Section 6). For that matter, in general, it is unnecessary to provide treatment on wash trading, if the crypto assets under investigation are established and traded in mainstream exchanges, even if these exchanges are unregulated.

In summary, the above statistics on $\beta_{r_{TT}}^{\ell}$ and $\beta_{\sigma_{TT}}^{\ell}$ support our Hypothesis 1, that treatment on wash trading significantly reduces the volatility of jumps (measured by liquidity diffusion), but only marginally affects the magnitude of jumps (measured by liquidity jump). Section 6 will provide additional empirical evidence that treatment on wash trading without liquidity-adjustment does not adequately improve the asset's responsiveness to the autoregressive models.

## 5. Liquidity-Adjusted ARMA-GARCH/EGARCH Models

In this subsection we briefly review liquidity-adjusted ARMA-GARCH/EGARCH construct for the modeling of univariate condition return by Deng and Zhou (2024).[6] We apply the standard ARMA($p,q$)-GARCH/EGARCH(1,1) model on both $r_t$ and $r_t^{\ell}$ (with and without treatment on wash trading) for each asset to estimate its conditional returns and variances for the in-sample observations (from April 27, 2020 to February 8, 2024 for a total of 1,383 trading days) with a 365-day rolling window.[7] The ARMA-GARCH/EGARCH constructs are given as follows:

$$ARMA(p,q): r_t = \delta + \sum_{i=1}^{p} \phi_i r_{t-i} - \sum_{j=1}^{q} \theta_j \epsilon_{t-j} + \epsilon_t \qquad (3a)$$

$$GARCH(1,1): \sigma_t^2 = \omega + a\epsilon_{t-1}^2 + b\sigma_{t-1}^2 \qquad (3b)$$

$$EGARCH(1,1): \log \sigma_t^2 = \omega + b g(Z_{t-1}) + a \log \sigma_{t-1}^2 \qquad (3c)$$

$$\text{where } g(Z_t) = \theta Z_t + \lambda(|Z_t| - E(|Z_t|)); \; Z_t \sim N(0,1)$$

In the ARMA stage, they use the Akaike Information Criterion (AIC) to select the best-fit values of $p$ and $q$ (p, q ≤ 4) for each rolling window (Equation 3a). In the GARCH/EGARCH stage, they adopt the (1,1) specification and use the AIC criteria to select between a GARCH (Equation 3b) specification and an EGARCH specification (Equation 3c) for each in-sample day *t*. They then apply Equation 3a to produce the one-period (*t+1*) forecasted conditional mean return vector ($\hat{\mu}_{t+1}^{arga}$) of the 10 assets for each rolling window for $r_t$, and $r_t^\ell$ with and without treatment.[8] We follow their modeling procedure in this paper.

---

[6] From this point on, the subscript *t* refers to a point in time with a daily interval, i.e., day *t*.
[7] Deng and Zhou (2024) establish that $r_t$ and $r_t^{\ell}$, with and without treatment on wash trading, are stationary series through Adam-Fuller tests and therefore can be modeled by the autoregressive models.

## 6. Empirical Tests on Liquidity-Adjusted Autoregressive Models

In this section we provide empirical evidence that the liquidity-adjusted autoregressive models of Section 5 offer better predictability on conditional return than their traditional counterparts, through comparing the performance of LVMV portfolios against that of the TMV ones.

### 6.1 Benchmark Portfolios

We construct the benchmark portfolios following Deng and Zhou (2024):

Portfolio 1: An equal-weight portfolio with each asset assigned an equal weight of 10%.

Portfolio 2: A market (equilibrium) portfolio with each asset assigned a weight proportional to its market weight (in amount).

Portfolio 3: A liquidity-weight portfolio with treatment on wash trading, with each asset assigned a weight proportional to its daily liquidity jump $\left(\sim \beta_{r_t}^\ell\right)$, for the short-term investors that take advantage of low transaction costs.

Portfolio 4: It is the same as Portfolio 3 but without treatment on wash trading.

Portfolio 5: An inverse-liquidity-weight portfolio with treatment on wash trading, with each asset assigned a weight that is inversely proportional to its daily liquidity jump $\left(\sim 1/\beta_{r_t}^\ell\right)$, for the long-term investors that seek a liquidity premium.

Portfolio 6: It is the same as Portfolio 5 but without treatment on wash trading.

---

[8] For the rest of the paper, the "current timestamp" is end of day *t*, thus a variable with a *t* subscript is "realized" (either a direct observation or a calculated value from direct observations), while a variable with a *t+1* subscript is a one-period forecasted value. We also use accent mark "$\bar{v}$" to represent a mean variable, accent mark "$\hat{v}$" to represent a forecasted variable, and no accent mark "$v$" to represent a realized variable.

All the benchmark portfolios do not include the risk-free asset USDT (zero weight).

**6.2 Standard MV Portfolios**

Following Deng and Zhou (2024), we then construct three MV portfolios: traditional (TMV), liquidity adjusted and with treatment on wash trading (LAMV), and liquidity-adjusted without treatment on wash trading (LAMV). The standard daily-optimized MV in a time-series construct can be analytically expressed as the following quadratic programming problem with constraints:

$$\max_{W_t} \left( \bar{\mu}_t W_t - \frac{\lambda_t}{2} W_t^H \bar{\Sigma}_t W_t \right); H \text{ is Transpose} \tag{4}$$

*subject to:*

$$\sum_i^N w_t^i = 1; \ i = USDT, ADA, BNB, BTC, ETC, ETH, LINK, LTC, MATIC, XMR, XRP; \ N = 11$$

$w_t^i \geq 0$; *long-only*

$w_t^{USDT} \leq 1$

$w_t^i \leq 0.300$ (3 × *equal weight*); $i \neq USDT$

*where:*

$$\lambda_t = \frac{r_{t_{mkt}}^P - r_t^{rf}}{\sigma_{t\,mkt}^2} = \frac{r_{t_{mkt}}^P - r_t^{USDT}}{\sigma_{t\,mkt}^2} = \frac{r_{t_{mkt}}^P}{\sigma_{t\,mkt}^2}$$

$r_{t_{mkt}}^P$ *is the return of the market or equilibrium portfolio on day t*, $\sigma_{t\,mkt}^2$ *is its variance of the rolling window;*
$r_t^{rf}$ *and* $r_t^{USDT}$ *are the returns of a risk-free asset and USDT, respectively. USDT is regarded as risk-free with 0 return*

In the standard MV construct of Equation 4, $\bar{\mu}_t$ is the mean (row) return vector of the ten-asset portfolio over a 365-day rolling window ending on day *t*, and $\bar{\Sigma}_t$ is the covariance matrix of daily returns of the constituent assets in that rolling window. Both $\bar{\mu}_t$ and $\bar{\Sigma}_t$ are realized and derived from available information up to day *t*. In addition, $W_t$ is the portfolio (column) weight vector to be optimized for day *t*. The daily MV portfolios are:

Portfolio 7: A TMV portfolio; $\bar{\mu}_t$ is the mean vector of $r_t$'s over the rolling window ending on day *t*, or $\bar{\mu}_{r_t}$; $\bar{\Sigma}_t$ is the covariance matrix of $r_t$'s for the rolling window, or $\bar{\Sigma}_{r_t}$.

Portfolio 8: A LAMV portfolio with treatment on wash trading when constructing $r_t^\ell$'s; $\bar{\mu}_t$ is the mean vector of $r_t^\ell$'s over the rolling window ending on day $t$, or $\bar{\mu}_{r_t^\ell}$; $\bar{\Sigma}_t$ is the covariance matrix of $r_t^\ell$'s for the rolling window, or $\bar{\Sigma}_{r_t^\ell}$.

Portfolio 9: It is the same as Portfolio 8 but without treatment on wash trading.

### 6.3 ARMA-GARCH/EGARCH-enhanced MV Portfolios

We follow Deng and Zhou (2024) to construct three additional ARMA-GARCH/EGARCH-enhanced MV portfolios with forecasted daily return vector. We rewrite Equation 4 by retaining $\bar{\Sigma}_t$ and replacing $\bar{\mu}_t$ with the ARMA-GARCH/EGARCH forecasted return vector, $\hat{\mu}_{t+1}^{arga}$. The portfolios are constructed as:

$$\max_{W_t} \left( \hat{\mu}_{t+1}^{arga} W_t - \frac{\lambda_t}{2} W_t^H \bar{\Sigma}_t W_t \right) \tag{5}$$

All the constraints for Equation 5 are the same as those for Equation 4. The proposed ARMA-GARCH/EGARCH-enhanced MV portfolios are:

Portfolio 10: An ARMA-GARCH/EGARCH-enhanced TMV portfolio; $\hat{\mu}_{t+1}^{arga}$ is the return vector of ARMA-GARCH/EGARCH forecasted $r_t$ values for day $t+1$, $\hat{\mu}_{r_{t+1}}^{arga}$; $\bar{\Sigma}_t$ is the covariance matrix of $r_t$'s for the rolling window, or $\bar{\Sigma}_{r_t}$.

Portfolio 11: An ARMA-GARCH/EGARCH-enhanced LAMV portfolio with treatment on wash trades; $\hat{\mu}_{t+1}^{arga}$ is the return vector of ARMA-GARCH/EGARCH forecasted $r_t^\ell$ values for day $t+1$, $\hat{\mu}_{r_{t+1}^\ell}^{arga}$; $\bar{\Sigma}_t$ is the covariance matrix of $r_t^\ell$'s for the rolling window, or $\bar{\Sigma}_{r_t^\ell}$.

Portfolio 12: It is the same as Portfolio 11 but without treatment on wash trades.

### 6.4 Portfolio Descriptive Statistics and Discussions

We use the annualized Sharpe Ratio ($SR_a$) to compare the performance of the portfolios:

$$SR_a = \frac{r_a^P - r_a^{rf}}{\sigma_a^P} = \frac{r_a^P - r_a^{USDT}}{\sigma_a^P} = \frac{r_a^P}{\sigma_a^P} \tag{6}$$

Where:
1. $r_a^P, \sigma_a^P$ are the annualized realized regular daily portfolio return and standard deviation.

Table 3 captures the maximum daily portfolio return (Panel A), the maximum daily portfolio volatility (Panel B), and the Sharpe Ratios of 12 portfolios. The benchmark portfolios (Portfolios 1 to 6) are on the top. The six MV portfolios (Portfolios 7 to 12) are arranged in such: the TMV portfolios with incremental forecast enhancement are listed on the left (Portfolios 7, 10), while their corresponding LAMV portfolios with treatment on wash trading are shown in the middle (Portfolios 8, 11), and without treatment on wash trading on the right (Portfolios 9, 12). That way, it is easy to observe the progress in portfolio return and volatility after applying each enhancement vertically within the TMV and LAMV portfolios, and at the same time conveniently to compare the difference in maximum portfolio return, volatility, and Sharpe Ratio between the TMV and LAMV portfolios after applying forecast-enhanced methodology horizontally.

From Panel A of Table 3, the maximum daily return of the standard TMV Portfolio 7 at 134.94% is much higher than that of the standard LAMV Portfolios 8 and 9 with and without the treatment on wash trading at 36.82% and 35.18%, respectively. This indicates that the asset-level liquidity adjustment greatly reduces the portfolio-level jumps (which is beyond the scope of this paper and will be addressed in a subsequent study), while treatment on wash trading has no impact on portfolio-level jumps, which is consistent with our earlier argument that the treatment on wash trading does not reduce the asset-level liquidity jump $\beta_{r_t}^{\ell}$ in a significant way.

The maximum daily return of the TMV Portfolio 10 with forecast enhancement drops to a lower level at 72.96%. For the LAMV portfolios, the maximum daily return of the forecast-enhanced portfolios (Portfolios 11 and 12) has no discontinuity from the standard portfolios (Portfolios 8 and 9) at 38.51% and 42.83% with and without the treatment. For the TMV portfolios, the result indicates that the ARMA-GARCH/EGARCH models try to "smoothen" the asset-level jumps with "extra effort," and when such jumps are too severe (extreme liquidity) the correction seems not adequate, rendering the autoregressive models less effective. However, for the LAMV portfolios, jumps are essentially removed and the (liquidity-adjusted) return can be modeled by autoregressive models effectively again, and treatment on wash trading has limited incremental effect on jump reduction. This evidence provides another direct empirical support to Deng and Zhou (2024) and actually relaxes their requirement of treatment on wash trading.

Panel B of Table 3 captures the maximum daily portfolio volatility for all the 12 portfolios. The maximum portfolio volatility exhibits the same pattern of changes across the incrementally enhanced TMV and LAMV portfolios (vertically), and similar values between the TMV and LAMV portfolios with the same incremental enhancement (horizontally). Panel C of Table 3 summarizes the performance of all 12 portfolios in terms of Sharpe Ratio (SR). The LAMV portfolios demonstrate a clear incremental improvement with forecast enhancement, while the TMV portfolios exhibit a performance deterioration. The standard TMV Portfolio 7 has a SR of 1.35, which drops drastically to 0.75 for the forecast enhanced TMV Portfolio 9. This result again supports Deng and Zhou (2024) that without reduction of jumps, the forecast of ARMA-GARCH/EGARCH on $r_{t+1}$ is actually less accurate than just averaging $r_t$'s in a rolling window. On the other hand, the LAMV portfolios demonstrates that forecast enhancement improves their performance. With treatment on wash trading, the SR of the standard LAMV Portfolio 8 from 0.96 is greatly improved to 1.45 for the forecast enhanced LAMV Portfolio 11. Even without treatment, the SR has also greatly improved from 1.41 of the standard LAMV Portfolio 9 to 1.81 for the forecast enhanced LAMV Portfolio 12. These results again indicate that the ARMA-GARCH/EGARCH model is highly effective in modeling the liquidity-adjusted return, with or without treatment on wash trading, and provide another direct empirical support to Deng and Zhou (2024) in relaxing their requirement of treatment on wash trading.

We notice that the standard LAMV portfolio with treatment on wash trading (Portfolio 8) has a lower SR (0.96) than that from the standard TMV Portfolio 7 (SR=1.35) and the standard LAMV Portfolio 9 without treatment on wash trading (SR=1.41). This seems to suggest that treatment on wash trading may have removed legitimate high-volume trades that are not initiated by manipulative traders with malicious intent, and therefore negatively impacted portfolio performance. As such, it may be better not to treat wash trading without having the ability to actually identify the real wash trades.

## 7. Conclusions

In this paper, we propose that the liquidity of an asset can be divided into two distinctive yet complementary components: liquidity jump and liquidity diffusion. The liquidity jump is defined as the ratio of regular return and liquidity-adjusted return that measures the magnitude of aggregated price jump in a given day, while the liquidity diffusion is the ratio between regular volatility and liquidity-adjusted volatility that reflects the intraday volatility of the aggregated daily jumps. The liquidity-adjusted return and volatility are developed in Deng and Zhou (2024).

We investigate the effect of treatment on wash trading (Deng and Zhou, 2024) in the effort of combating the wash trades embedded in crypto trading (Cong et al., 2023). We find that the liquidity diffusion has a higher correlation with wash trading than the liquidity jump, and that a combination of high liquidity jump and high liquidity diffusion is the most reliable indicator for wash trading. An explanation is that for crypto assets with high market-cap and long trading history that trade in established exchanges, the manipulative traders do not conduct a small number of very large-volume trades as it is unlikely for them to maintain the long-term upward price movement. Rather, they engage in high-frequency, large number of relatively small-volume momentum trades that award them with small, frequent, and sustainable short-term gains. These trades in turn increase liquidity volatility, resulting in higher levels of liquidity diffusion. The distribution of the liquidity jump and liquidity diffusion indicates that wash trading is uncommon among high market-cap crypto assets that trade in established exchanges. On the other hand, the majority of very large-volume trades that induce higher level of liquidity jump may actually reflect legitimate high demand with unregulated trading. We demonstrate that treatment of washing trading significantly reduces the liquidity diffusion, but only reduces the liquidity jump to a level that is inadequate to restore the asset's responsiveness to the autoregressive models.

We use the forecasted daily liquidity-adjusted return from the ARMA-GARCH/EGARCH models (Deng and Zhou, 2024) as inputs to the LAMV constructs for portfolio optimization. For comparison, we duplicate the procedure with the forecasted daily regular (non-liquidity-adjusted) return being used as inputs to the TMV constructs. We find that the ARMA-GARCH/EGARCH models are highly effective in modeling the liquidity-adjusted return as we observe a clear advantage for the LAMV over the TMV in portfolio optimization, with and without treatment on wash trading. We also notice that the standard LAMV portfolio with treatment on wash trading actually has an inferior performance to that without the treatment, suggesting that the treatment may have removed a greater number of legitimate high-volume trades than actual wash trades, as there are only a negligible number of them to begin with. Therefore, in general, it is unnecessary to treat wash trading in modeling established crypto assets that trade in mainstream exchanges, even if these exchanges are unregulated.

In summary, by studying the behavior of the proposed and liquidity diffusion, we establish that the liquidity adjustment proposed by Deng and Zhou (2024) reduces the level of liquidity jump adequately in restoring the autoregressive properties to the (liquidity-adjusted) return and volatility of assets with extreme liquidity, while treatment on wash trading is not needed, as although it reduces the level of liquidity diffusion, it does not reduce the level of liquidity jump to adequately restore the effectiveness of the autoregressive models. To some extent, treatment on wash trading may actually deteriorate the effectiveness of modeling as it removes legitimate high-volume trades. Our liquidity jump-diffusion model provides a viable and robust alternative for modeling asset-level liquidity and its components of liquidity jump and liquidity diffusion of crypto assets and can be utilized to model other asset classes with high liquidity risk. Furthermore, we provide the investors a simple and practical investment strategy for crypto assets, that they may consider adding established crypto assets traded in mainstream exchanges as an alternative asset class to their investment portfolios, with the appropriate modeling technique of liquidity adjustment to the asset return and volatility outlined in this paper.

**Disclosure Statement**

The authors declare that they have no known competing financial interests or personal relationships that could have appeared to influence the work reported in this paper.

# Table 1 - Descriptive Statistics of Individual Asset $\beta^{\ell}_{r_{TT}}$

This table reports descriptive statistics of liquidity jump ($\beta^{\ell}_{r_{TT}}$) for each crypto asset over the entire sample period. The maximum value of $\beta^{\ell}_{r_{TT}}$ is capped at 10. All ten crypto assets and the portfolio are measured with their trading pairs with Tether or USDT, a "stable coin" pegged to the US dollar, which is regarded as the "risk-free" asset in portfolios with a 0% interest rate in terms of their market values.

| Panel A (wash trades removed) | | | | | liquidity jump ($\beta^{\ell}_{r_{TT}}$) | | | | | |
|---|---|---|---|---|---|---|---|---|---|---|
| ticker | ADA | BNB | BTC | ETC | ETH | LINK | LTC | MATIC | XMR | XRP |
| count | 1749 | 1749 | 1749 | 1749 | 1749 | 1749 | 1749 | 1749 | 1749 | 1749 |
| mean | 1.57 | 1.55 | 1.19 | 1.53 | 1.36 | 1.38 | 1.51 | 1.41 | 1.47 | 1.34 |
| std | 2.16 | 2.09 | 1.68 | 2.08 | 1.88 | 1.87 | 2.00 | 1.96 | 2.12 | 1.96 |
| min | 0.00 | 0.00 | 0.00 | 0.00 | 0.01 | 0.00 | 0.00 | 0.00 | 0.00 | 0.00 |
| 25% | 0.51 | 0.54 | 0.48 | 0.47 | 0.56 | 0.49 | 0.51 | 0.48 | 0.35 | 0.47 |
| 50% (median) | 0.84 | 0.91 | 0.73 | 0.84 | 0.81 | 0.83 | 0.87 | 0.83 | 0.78 | 0.75 |
| 75% | 1.48 | 1.47 | 1.15 | 1.53 | 1.27 | 1.36 | 1.53 | 1.37 | 1.43 | 1.23 |
| max | 10.00 | 10.00 | 10.00 | 10.00 | 10.00 | 10.00 | 10.00 | 10.00 | 10.00 | 10.00 |
| highest days (= max) | 64 | 62 | 30 | 55 | 52 | 38 | 53 | 47 | 61 | 51 |
| % of total days | 3.66% | 3.54% | 1.72% | 3.14% | 2.97% | 2.17% | 3.03% | 2.69% | 3.49% | 2.92% |
| weight in beta | 23.25% | 22.83% | 14.40% | 20.55% | 21.92% | 15.73% | 20.09% | 19.13% | 23.81% | 21.76% |
| highest days (>= mean) | 401 | 416 | 408 | 437 | 386 | 425 | 443 | 421 | 429 | 385 |
| % of total days | 22.93% | 23.79% | 23.33% | 24.99% | 22.07% | 24.30% | 25.33% | 24.07% | 24.53% | 22.01% |
| highest days (>= 1) | 730 | 764 | 534 | 724 | 620 | 677 | 745 | 674 | 658 | 591 |
| % of total days | 41.74% | 43.68% | 30.53% | 41.40% | 35.45% | 38.71% | 42.60% | 38.54% | 37.62% | 33.79% |
| lowest days (<= 0.10) | 63 | 80 | 75 | 69 | 61 | 89 | 60 | 96 | 128 | 62 |
| % of total days | 3.60% | 4.57% | 4.29% | 3.95% | 3.49% | 5.09% | 3.43% | 5.49% | 7.32% | 3.54% |

| Panel B (wash trades retained) | | | | | liquidity jump ($\beta^{\ell}_{r_{TT}}$) | | | | | |
|---|---|---|---|---|---|---|---|---|---|---|
| ticker | ADA | BNB | BTC | ETC | ETH | LINK | LTC | MATIC | XMR | XRP |
| count | 1749 | 1749 | 1749 | 1749 | 1749 | 1749 | 1749 | 1749 | 1749 | 1749 |
| mean | 2.02 | 1.78 | 1.49 | 1.99 | 1.60 | 1.97 | 1.99 | 1.88 | 1.98 | 1.59 |
| std | 2.51 | 2.30 | 1.95 | 2.49 | 2.00 | 2.50 | 2.53 | 2.46 | 2.66 | 2.14 |
| min | 0.00 | 0.00 | 0.00 | 0.00 | 0.01 | 0.00 | 0.00 | 0.00 | 0.00 | 0.00 |
| 25% | 0.58 | 0.52 | 0.56 | 0.52 | 0.59 | 0.56 | 0.49 | 0.48 | 0.41 | 0.48 |
| 50% (median) | 1.06 | 1.00 | 0.91 | 1.03 | 0.98 | 1.00 | 1.01 | 1.00 | 0.90 | 0.85 |
| 75% | 2.15 | 1.85 | 1.43 | 2.22 | 1.64 | 2.05 | 2.16 | 1.92 | 2.11 | 1.61 |
| max | 10.00 | 10.00 | 10.00 | 10.00 | 10.00 | 10.00 | 10.00 | 10.00 | 10.00 | 10.00 |
| highest days (= max) | 91 | 77 | 44 | 90 | 52 | 92 | 78 | 89 | 110 | 53 |
| % of total days | 5.20% | 4.40% | 2.52% | 5.15% | 2.97% | 5.26% | 4.46% | 5.09% | 6.29% | 3.03% |
| weight in beta | 25.78% | 24.77% | 16.85% | 25.90% | 18.61% | 26.67% | 22.42% | 27.12% | 31.81% | 19.10% |
| highest days (>= mean) | 463 | 459 | 412 | 472 | 451 | 458 | 479 | 448 | 453 | 443 |
| % of total days | 26.47% | 26.24% | 23.56% | 26.99% | 25.79% | 26.19% | 27.39% | 25.61% | 25.90% | 25.33% |
| highest days (>= 1) | 921 | 874 | 772 | 906 | 854 | 876 | 880 | 870 | 817 | 731 |
| % of total days | 52.66% | 49.97% | 44.14% | 51.80% | 48.83% | 50.09% | 50.31% | 49.74% | 46.71% | 41.80% |
| lowest days (<= 0.10) | 67 | 89 | 58 | 65 | 62 | 72 | 79 | 108 | 112 | 73 |
| % of total days | 3.83% | 5.09% | 3.32% | 3.72% | 3.54% | 4.12% | 4.52% | 6.17% | 6.40% | 4.17% |

## Table 2 - Descriptive Statistics of Individual Asset $\beta^{\ell}_{\sigma_{TT}}$

This table reports descriptive statistics of liquidity diffusion ($\beta^{\ell}_{\sigma_{TT}}$) for each crypto asset over the entire sample period. The maximum value of $\beta^{\ell}_{\sigma_{TT}}$ is capped at 10. All ten crypto assets and the portfolio are measured with their trading pairs with Tether or USDT, a "stable coin" pegged to the US dollar, which is regarded as the "risk-free" asset in portfolios with a 0% interest rate in terms of their market values.

| Panel A (wash trades removed) | | | | liquidity diffusion ($\beta^{\ell}_{\sigma_{TT}}$) | | | | | | |
|---|---|---|---|---|---|---|---|---|---|---|
| ticker | ADA | BNB | BTC | ETC | ETH | LINK | LTC | MATIC | XMR | XRP |
| count | 1749 | 1749 | 1749 | 1749 | 1749 | 1749 | 1749 | 1749 | 1749 | 1749 |
| mean | 0.82 | 0.73 | 0.66 | 0.87 | 0.70 | 0.81 | 0.81 | 0.80 | 0.87 | 0.72 |
| std | 0.12 | 0.07 | 0.05 | 0.12 | 0.04 | 0.11 | 0.11 | 0.11 | 0.10 | 0.07 |
| min | 0.39 | 0.09 | 0.36 | 0.40 | 0.43 | 0.46 | 0.46 | 0.42 | 0.43 | 0.38 |
| 25% | 0.72 | 0.68 | 0.64 | 0.78 | 0.68 | 0.72 | 0.73 | 0.72 | 0.82 | 0.67 |
| 50% (median) | 0.79 | 0.72 | 0.67 | 0.86 | 0.70 | 0.78 | 0.78 | 0.76 | 0.88 | 0.70 |
| 75% | 0.90 | 0.77 | 0.69 | 0.94 | 0.73 | 0.88 | 0.86 | 0.87 | 0.93 | 0.75 |
| max | 1.30 | 1.06 | 0.85 | 1.53 | 0.92 | 2.07 | 1.48 | 1.38 | 1.55 | 1.60 |
| highest days (= max) | 0 | 0 | 0 | 0 | 0 | 0 | 0 | 0 | 0 | 0 |
| % of total days | 0.00% | 0.00% | 0.00% | 0.00% | 0.00% | 0.00% | 0.00% | 0.00% | 0.00% | 0.00% |
| weight in beta | 0.00% | 0.00% | 0.00% | 0.00% | 0.00% | 0.00% | 0.00% | 0.00% | 0.00% | 0.00% |
| highest days (>= mean) | 756 | 776 | 991 | 807 | 877 | 739 | 688 | 713 | 948 | 695 |
| % of total days | 43.22% | 44.37% | 56.66% | 46.14% | 50.14% | 42.25% | 39.34% | 40.77% | 54.20% | 39.74% |
| highest days (>= 1) | 162 | 4 | 0 | 209 | 0 | 91 | 126 | 77 | 125 | 6 |
| % of total days | 9.26% | 0.23% | 0.00% | 11.95% | 0.00% | 5.20% | 7.20% | 4.40% | 7.15% | 0.34% |
| lowest days (<= 0.10) | 0 | 1 | 0 | 0 | 0 | 0 | 0 | 0 | 0 | 0 |
| % of total days | 0.00% | 0.06% | 0.00% | 0.00% | 0.00% | 0.00% | 0.00% | 0.00% | 0.00% | 0.00% |

| Panel B (wash trades retained) | | | | liquidity diffusion ($\beta^{\ell}_{\sigma_{TT}}$) | | | | | | |
|---|---|---|---|---|---|---|---|---|---|---|
| ticker | ADA | BNB | BTC | ETC | ETH | LINK | LTC | MATIC | XMR | XRP |
| count | 1749 | 1749 | 1749 | 1749 | 1749 | 1749 | 1749 | 1749 | 1749 | 1749 |
| mean | 1.33 | 0.90 | 0.82 | 1.40 | 0.87 | 1.38 | 1.11 | 1.14 | 1.49 | 0.92 |
| std | 1.29 | 0.15 | 0.07 | 0.67 | 0.08 | 1.01 | 0.68 | 0.85 | 1.34 | 0.33 |
| min | 0.55 | 0.58 | 0.43 | 0.55 | 0.52 | 0.56 | 0.65 | 0.48 | 0.60 | 0.57 |
| 25% | 0.87 | 0.81 | 0.78 | 0.98 | 0.82 | 0.88 | 0.88 | 0.86 | 0.98 | 0.82 |
| 50% (median) | 0.98 | 0.86 | 0.82 | 1.18 | 0.86 | 1.01 | 0.96 | 0.93 | 1.12 | 0.87 |
| 75% | 1.27 | 0.93 | 0.86 | 1.59 | 0.91 | 1.41 | 1.11 | 1.11 | 1.39 | 0.96 |
| max | 10.00 | 2.53 | 1.15 | 8.89 | 1.45 | 10.00 | 10.00 | 10.00 | 10.00 | 10.00 |
| highest days (= max) | 18 | 0 | 0 | 0 | 0 | 3 | 2 | 7 | 19 | 1 |
| % of total days | 1.03% | 0.00% | 0.00% | 0.00% | 0.00% | 0.17% | 0.11% | 0.40% | 1.09% | 0.06% |
| weight in beta | 7.74% | 0.00% | 0.00% | 0.00% | 0.00% | 1.24% | 1.03% | 3.52% | 7.29% | 0.62% |
| highest days (>= mean) | 362 | 597 | 840 | 589 | 759 | 451 | 452 | 397 | 353 | 564 |
| % of total days | 20.70% | 34.13% | 48.03% | 33.68% | 43.40% | 25.79% | 25.84% | 22.70% | 20.18% | 32.25% |
| highest days (>= 1) | 821 | 261 | 18 | 1248 | 108 | 888 | 692 | 645 | 1262 | 320 |
| % of total days | 46.94% | 14.92% | 1.03% | 71.36% | 6.17% | 50.77% | 39.57% | 36.88% | 72.16% | 18.30% |
| lowest days (<= 0.10) | 0 | 0 | 0 | 0 | 0 | 0 | 0 | 0 | 0 | 0 |
| % of total days | 0.00% | 0.00% | 0.00% | 0.00% | 0.00% | 0.00% | 0.00% | 0.00% | 0.00% | 0.00% |

**Table 3 - Portfolio Performance: Max Return ($r^P_{t+1}$) and Volatility ($\sigma^P_{t+1}$) and Sharpe Ratio**

This table compares the maximum portfolio return ($r^P_{t+1}$), maximum portfolio volatility ($\sigma^P_{t+1}$) and Sharpe Ratio for all 12 portfolios, with and without treatment on wash trading.

| Panel A | Portfolio Return | | | | | |
|---|---|---|---|---|---|---|
| Portfolio Number | 1 | 2 | 3 | 4 | 5 | 6 |
| Portfolio Description | equ | mkt | blq - w/treatment | blq - w/o treatment | blq_inv - w/treatment | blq_inv - w/o treatment |
| max daily return | 77.78% | 85.77% | 144.18% | 68.51% | 46.29% | 70.45% |
| Portfolio Number | 7 | | 8 | | 9 | |
| Portfolio Description | MV_rr | | MV_rrlq - w/ treatment | | MV_rrlq - w/o treatment | |
| max daily return | 134.94% | | 36.82% | | 35.18% | |
| Portfolio Number | 10 | | 11 | | 12 | |
| Portfolio Description | MV_arga_rr | | MV_arga_rrlq - w/ treatment | | MV_arga_rrlq - w/o treatment | |
| max daily return | 72.96% | | 38.51% | | 42.83% | |

| Panel B | Portfolio Volatility | | | | | |
|---|---|---|---|---|---|---|
| Portfolio Number | 1 | 2 | 3 | 4 | 5 | 6 |
| Portfolio Description | equ | mkt | blq - w/treatment | blq - w/o treatment | blq_inv - w/treatment | blq_inv - w/o treatment |
| max daily volatility | 31.83% | 31.15% | 34.54% | 34.48% | 30.11% | 29.89% |
| Portfolio Number | 7 | | 8 | | 9 | |
| Portfolio Description | MV_rr | | MV_rrlq - w/ treatment | | MV_rrlq - w/o treatment | |
| max daily volatility | 41.09% | | 40.16% | | 40.73% | |
| Portfolio Number | 10 | | 11 | | 12 | |
| Portfolio Description | MV_arga_rr | | MV_arga_rrlq - w/ treatment | | MV_arga_rrlq - w/o treatment | |
| max daily volatility | 51.81% | | 51.80% | | 38.79% | |

| Panel C | Portfolio Sharpe Ratio | | | | | |
|---|---|---|---|---|---|---|
| Portfolio Number | 1 | 2 | 3 | 4 | 5 | 6 |
| Portfolio Description | equ | mkt | blq - w/treatment | blq - w/o treatment | blq_inv - w/treatment | blq_inv - w/o treatment |
| annualized Sharpe Ratio ($R_f$=0%) | 1.44 | 1.24 | 1.38 | 1.13 | 0.96 | 1.09 |
| Portfolio Number | 7 | | 8 | | 9 | |
| Portfolio Description | MV_rr | | MV_rrlq - w/ treatment | | MV_rrlq - w/o treatment | |
| annualized Sharpe Ratio ($R_f$=0%) | 1.35 | | 0.96 | | 1.41 | |
| Portfolio Number | 10 | | 11 | | 12 | |
| Portfolio Description | MV_arga_rr | | MV_arga_rrlq - w/ treatment | | MV_arga_rrlq - w/o treatment | |
| annualized Sharpe Ratio ($R_f$=0%) | 0.75 | | 1.45 | | 1.81 | |

# Figure 1 – Distribution of $\beta^{\ell}_{r_{TT}}$-$\beta^{\ell}_{\sigma_{TT}}$ with Treatment on Wash Trading

This figure provides the scatter plots of liquidity jump ($\beta^{\ell}_{r_{TT}}$) vs. liquidity diffusion ($\beta^{\ell}_{\sigma_{TT}}$) with treatment on wash trading for each individual asset over the entire sample period. The max values of $\beta^{\ell}_{r_{TT}}$ and $\beta^{\ell}_{\sigma_{TT}}$ are capped at 10.

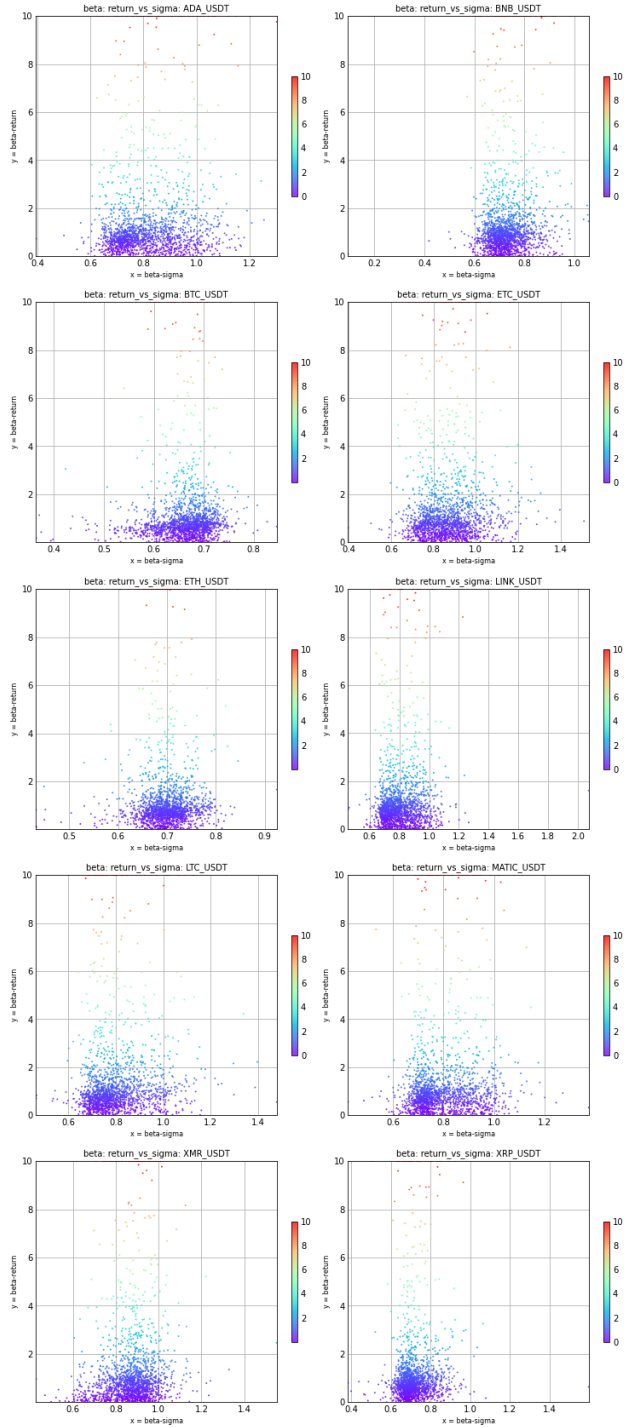

**Figure 2 – Distribution of $\beta^{\ell}_{r_{TT}}$-$\beta^{\ell}_{\sigma_{TT}}$ without Treatment on Wash Trading**

This figure provides the scatter plots of liquidity jump ($\beta^{\ell}_{r_{TT}}$) and liquidity diffusion ($\beta^{\ell}_{\sigma_{TT}}$) without treatment on wash trading for each individual asset over the entire sample period. The max values of $\beta^{\ell}_{r_{TT}}$ and $\beta^{\ell}_{\sigma_{TT}}$ are capped at 10.

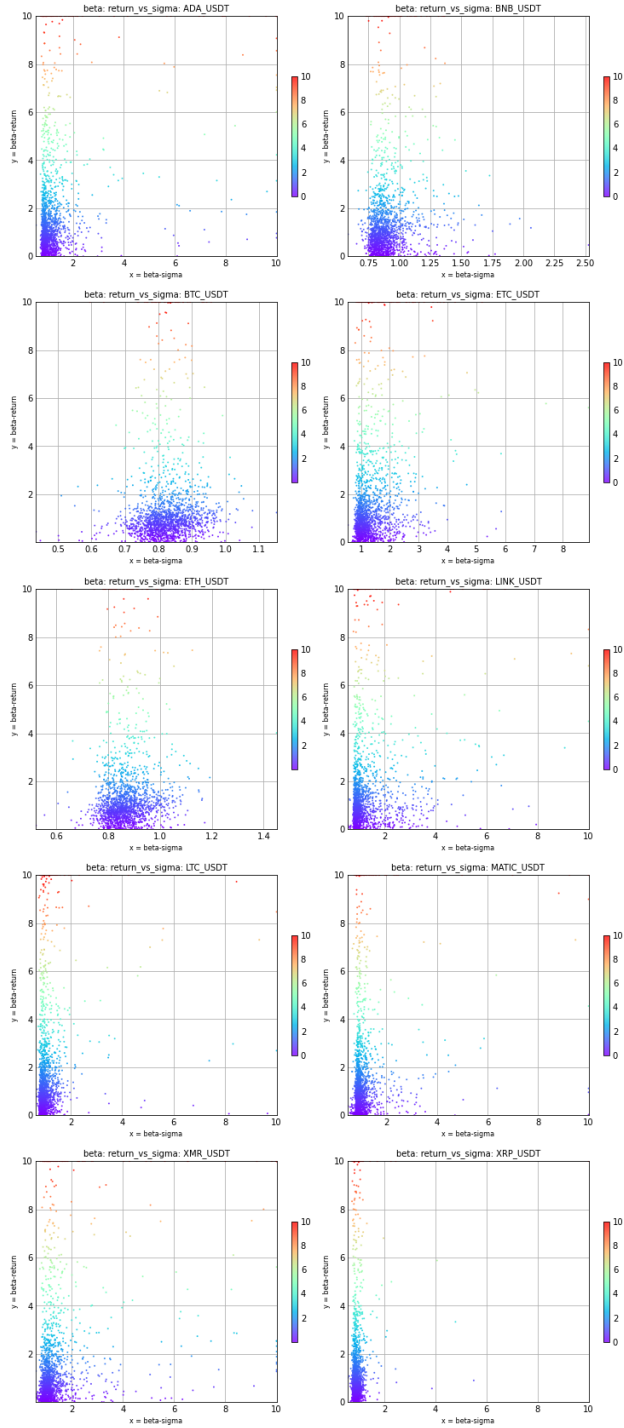

**Figure 3 – Distribution of $\beta^{\ell}_{r_{TT}}$ with Treatment on Wash Trading**

This figure provides the distribution of liquidity jump ($\beta^{\ell}_{r_{TT}}$) for each individual asset over the entire sample period. The maximum value of $\beta^{\ell}_{r_{TT}}$ is capped at 10. Column 1 presents the histograms, Column 2 shows the scatter plots of ($r_{TT}$ vs. $r^{\ell}_{TT}$) overlapping a straight line with coefficient of 1.0 ($\beta^{\ell}_{r_{TT}}$ =1.0 or equilibrium value), and Column 3 provides the 3D scatter plots of ($r_{TT}$-$r^{\ell}_{TT}$-$\beta^{\ell}_{r_{TT}}$) with its 2D projection ($r_{TT}$-$r^{\ell}_{TT}$).

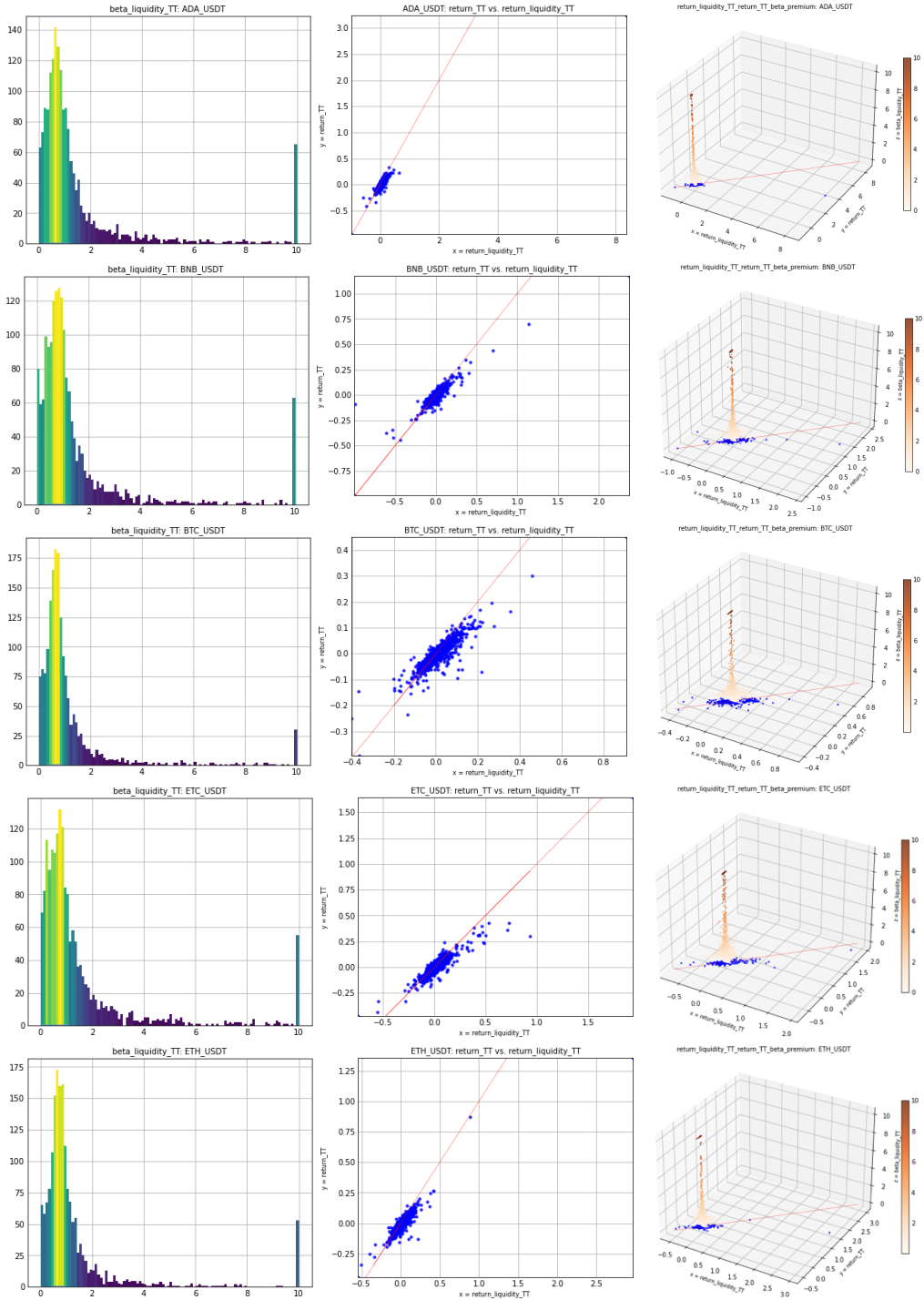

# Figure 3 – Distribution of $\beta^{\ell}_{r_{TT}}$ with Treatment on Wash Trading – Continued

This figure provides the distribution of liquidity jump ($\beta^{\ell}_{r_{TT}}$) for each individual asset over the entire sample period. The maximum value of $\beta^{\ell}_{r_{TT}}$ is capped at 10. Column 1 presents the histograms, Column 2 shows the scatter plots of ($r_{TT}$ vs. $r^{\ell}_{TT}$) overlapping a straight line with coefficient of 1.0 ($\beta^{\ell}_{r_{TT}}$ =1.0 or equilibrium value), and Column 3 provides the 3D distribution of return-volatility-liquidity ($r_{TT}$-$r^{\ell}_{TT}$-$\beta^{\ell}_{r_{TT}}$) with its 2D projection ($r_{TT}$-$r^{\ell}_{TT}$).

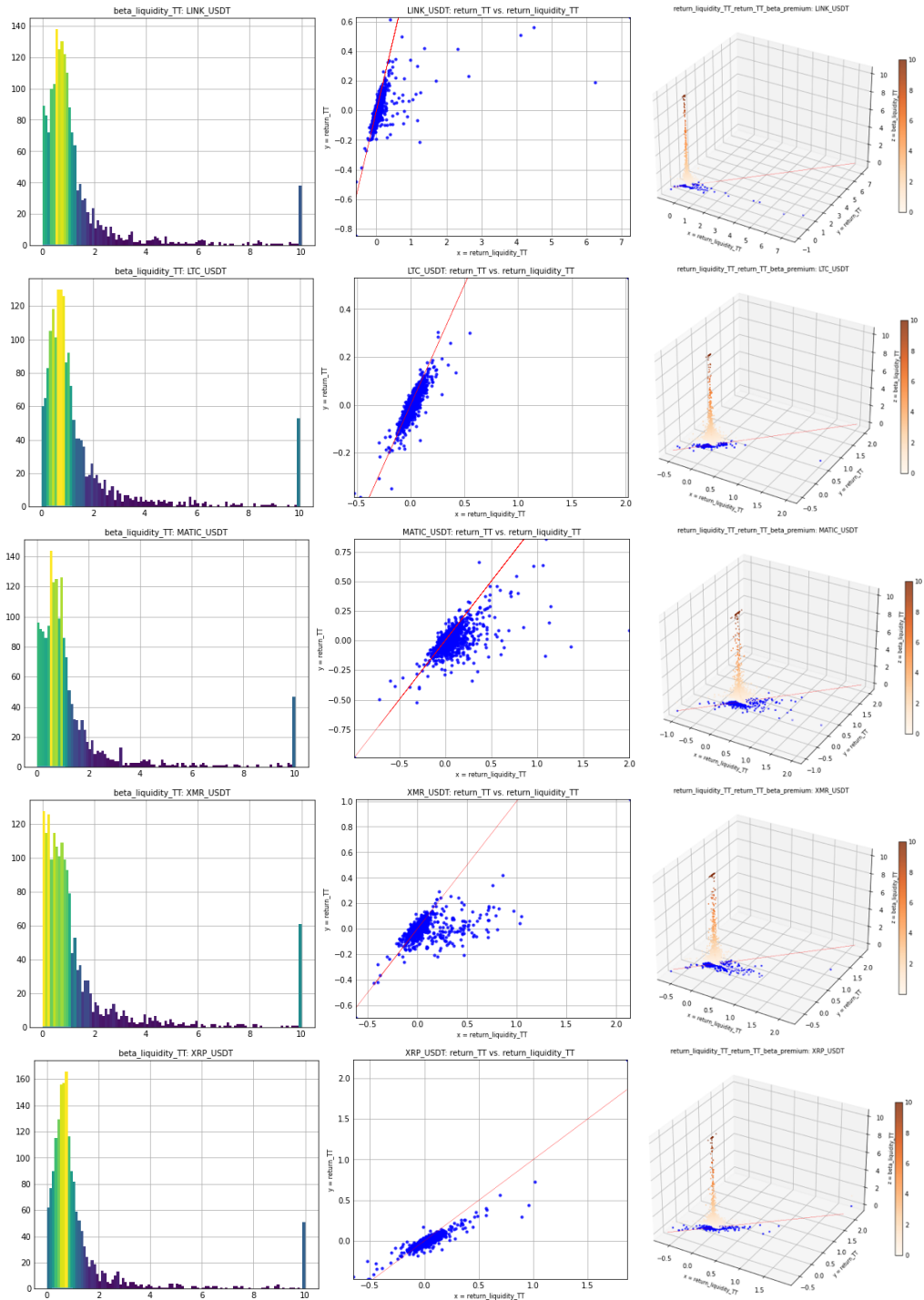

# Figure 4 – Distribution of $\beta^{\ell}_{r_{TT}}$ without Treatment on Wash Trading

This figure provides the distribution of liquidity jump ($\beta^{\ell}_{r_{TT}}$) for each individual asset over the entire sample period. The maximum value of $\beta^{\ell}_{r_{TT}}$ is capped at 10. Column 1 presents the histograms, Column 2 shows the scatter plots of ($r_{TT}$ vs. $r^{\ell}_{TT}$) overlapping a straight line with coefficient of 1.0 ($\beta^{\ell}_{r_{TT}}$ =1.0 or equilibrium value), and Column 3 provides the 3D scatter plots of ($r_{TT}$-$r^{\ell}_{TT}$-$\beta^{\ell}_{r_{TT}}$) with its 2D projection ($r_{TT}$-$r^{\ell}_{TT}$).

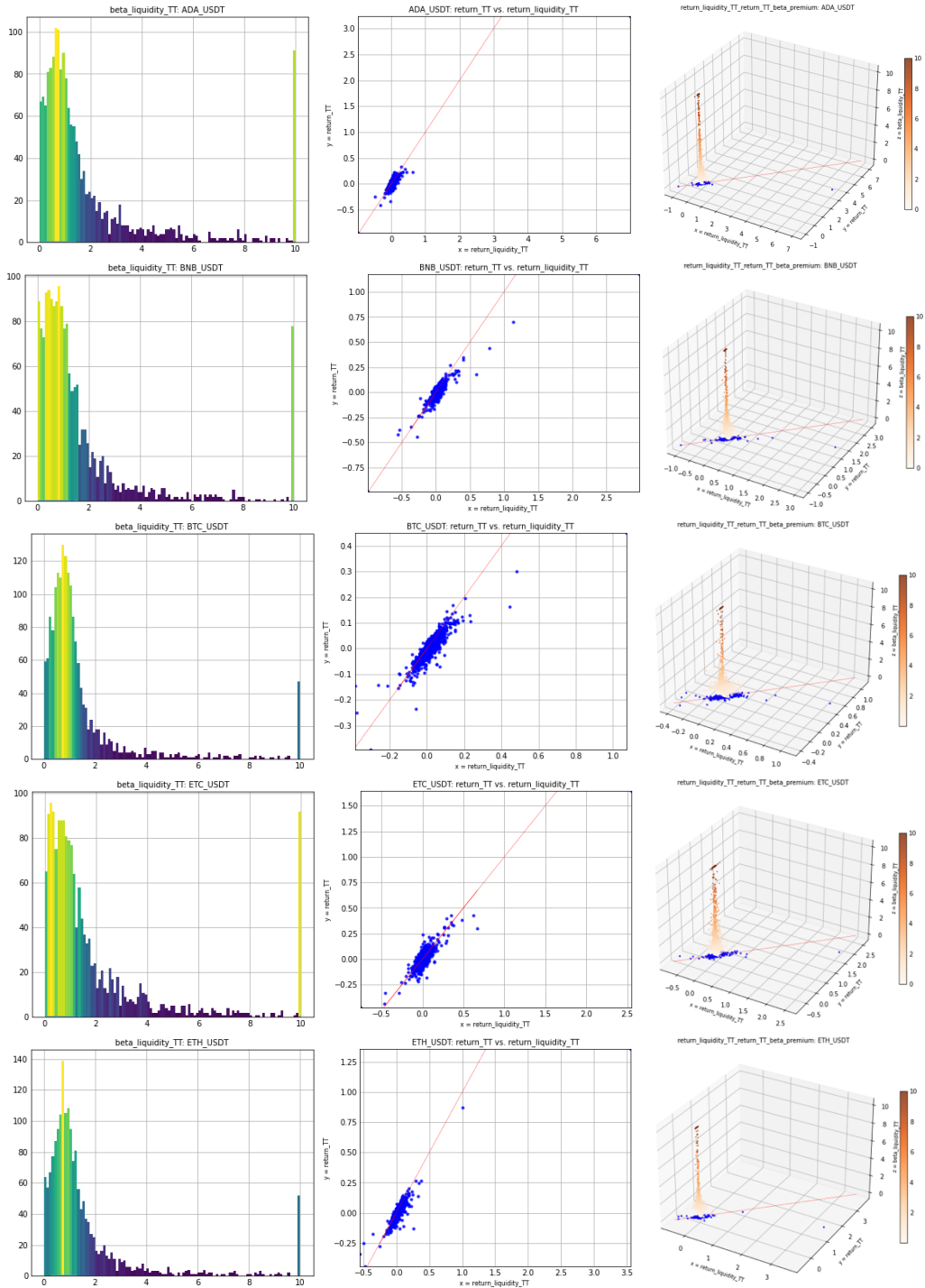

**Figure 4 – Distribution of $\beta^{\ell}_{r_{TT}}$ without Treatment on Wash Trading – Continued**

This figure provides the distribution of liquidity jump ($\beta^{\ell}_{r_{TT}}$) for each individual asset over the entire sample period. The maximum value of $\beta^{\ell}_{r_{TT}}$ is capped at 10. Column 1 presents the histograms, Column 2 shows the scatter plots of ($r_{TT}$ vs. $r^{\ell}_{TT}$) overlapping a straight line with coefficient of 1.0 ($\beta^{\ell}_{r_{TT}} =1.0$ or equilibrium value), and Column 3 provides the 3D distribution of return-volatility-liquidity ($r_{TT}$-$r^{\ell}_{TT}$-$\beta^{\ell}_{r_{TT}}$) with its 2D projection ($r_{TT}$-$r^{\ell}_{TT}$).

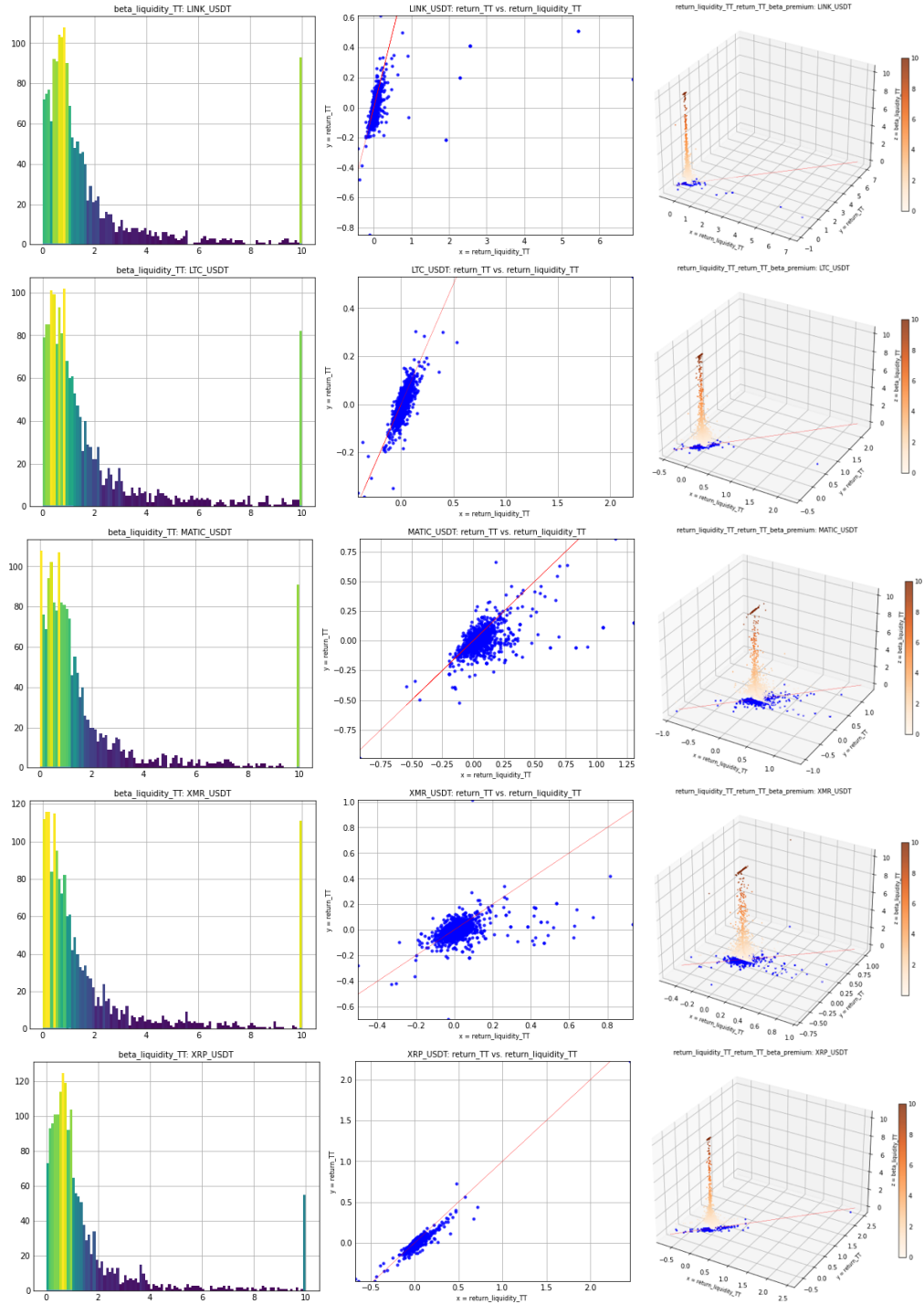

# Figure 5 – Distribution of $\beta^{\ell}_{\sigma_{TT}}$ with Treatment on Wash Trading

This figure provides the distribution of liquidity diffusion ($\beta^{\ell}_{\sigma_{TT}}$) for each individual asset over the entire sample period. Column 1 presents the histograms, Column 2 shows the scatter plots of ($\sigma_{TT}$ vs. $\sigma^{\ell}_{TT}$) overlapping a straight line with a coefficient of 1.0, and Column 3 provides the 3D scatter plots of $\sigma_{TT}$-$\sigma^{\ell}_{TT}$-$\beta^{\ell}_{\sigma_{TT}}$) with its 2D projection ($\sigma_{TT}$-$\sigma^{\ell}_{TT}$).

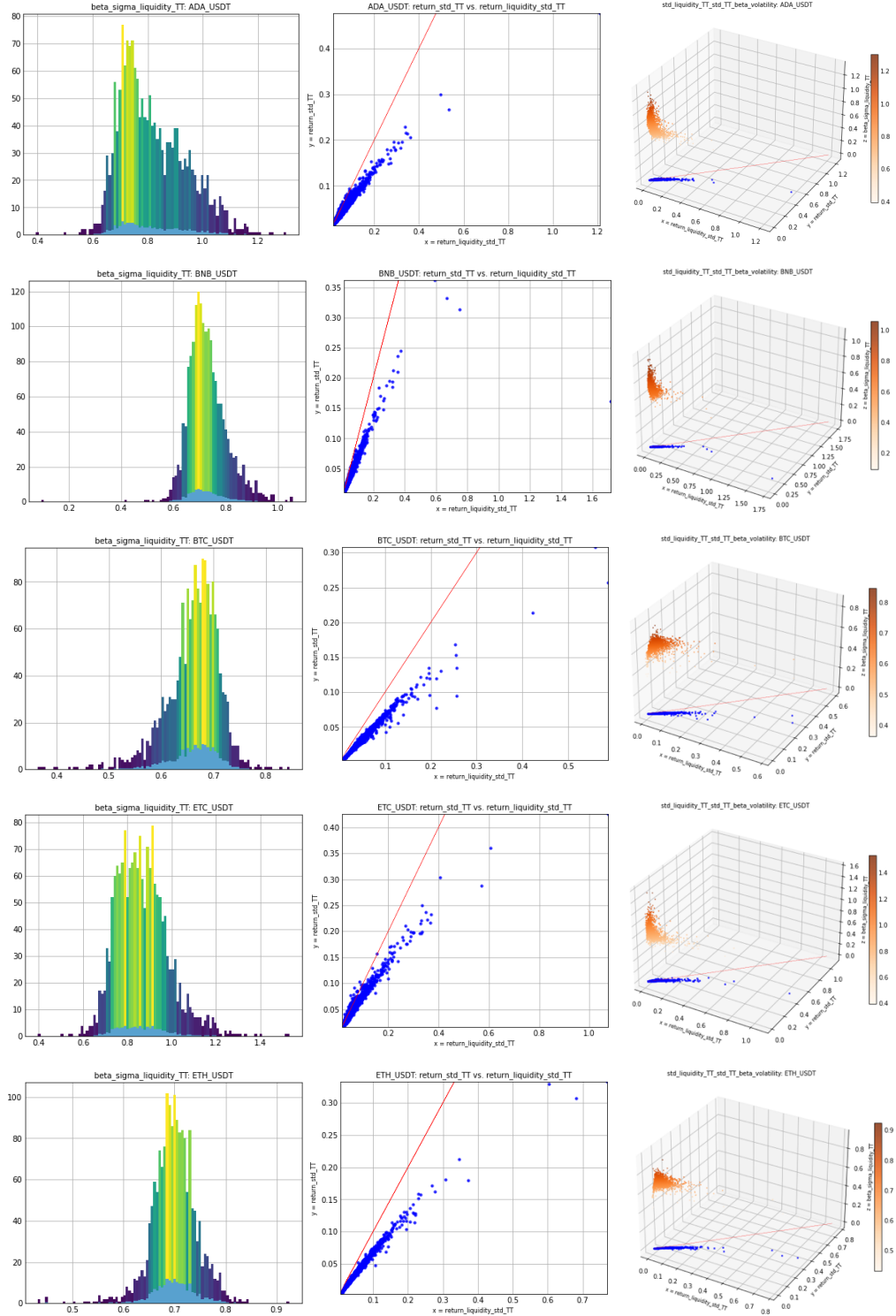

**Figure 5 – Distribution of $\beta^{\ell}_{\sigma_{TT}}$ with Treatment on Wash Trading – Continued**

This figure provides the distribution of liquidity diffusion ($\beta^{\ell}_{\sigma_{TT}}$) for each individual asset, over the entire sample period. Column 1 presents the histograms, Column 2 shows the scatter plots of ($\sigma_{TT}$ vs. $\sigma^{\ell}_{TT}$) overlapping a straight line with coefficient of 1.0, and Column 3 provides the 3D scatter plots of ($\sigma_{TT}$-$\sigma^{\ell}_{TT}$-$\beta^{\ell}_{\sigma_{TT}}$) with its 2D projection ($\sigma_{TT}$-$\sigma^{\ell}_{TT}$).

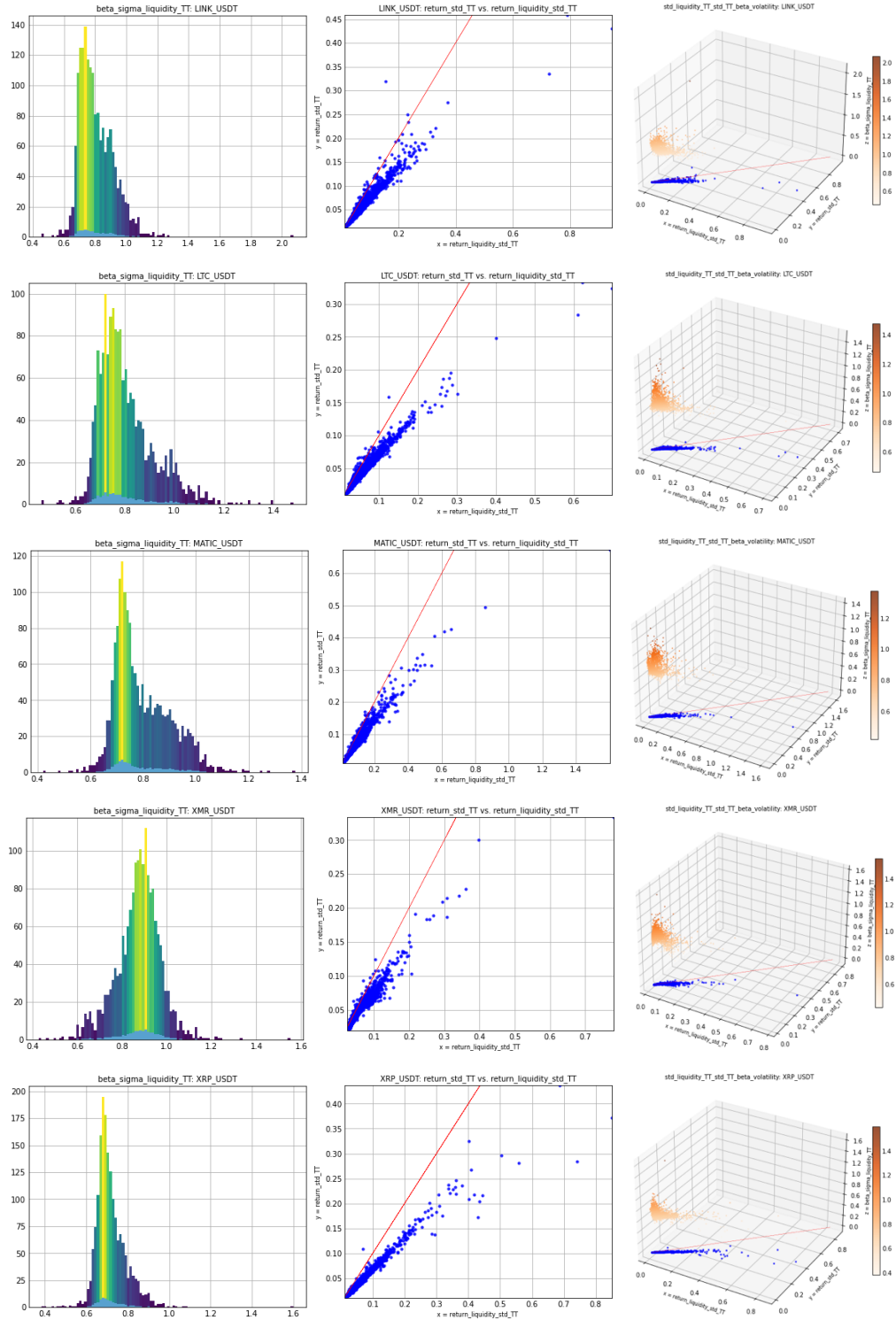

**Figure 6 – Distribution of $\beta^{\ell}_{\sigma_{TT}}$ without Treatment on Wash Trading**

This figure provides the distribution of liquidity diffusion ($\beta^{\ell}_{\sigma_{TT}}$) for each individual asset over the entire sample period. Column 1 presents the histograms, Column 2 shows the scatter plots of ($\sigma_{TT}$ vs. $\sigma^{\ell}_{TT}$) overlapping a straight line with coefficient of 1.0, and Column 3 provides the 3D scatter plots of $\sigma_{TT}$-$\sigma^{\ell}_{TT}$-$\beta^{\ell}_{\sigma_{TT}}$) with its 2D projection ($\sigma_{TT}$-$\sigma^{\ell}_{TT}$).

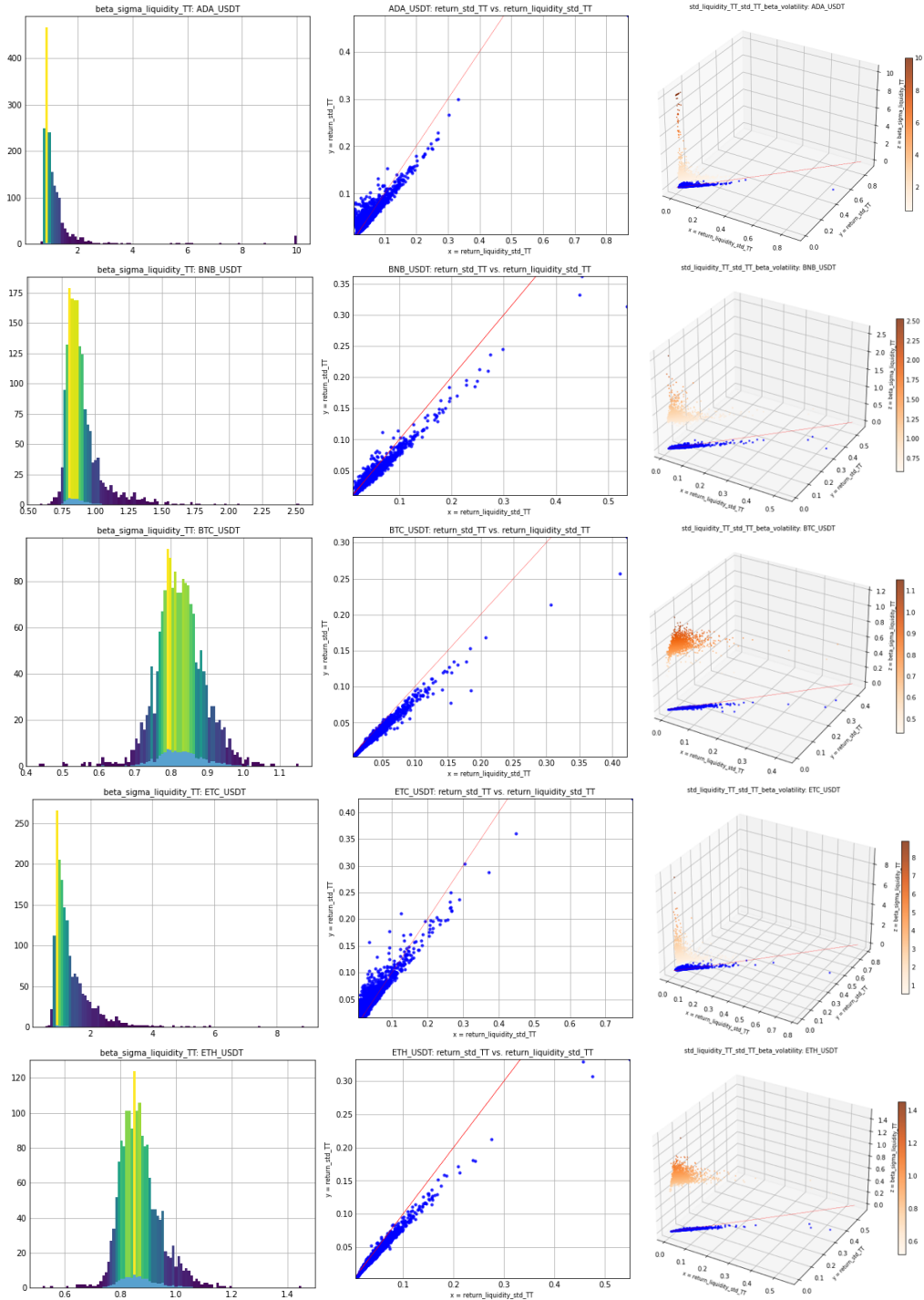

**Figure 6 – Distribution of $\beta^{\ell}_{\sigma_{TT}}$ without Treatment on Wash Trading– Continued**

This figure provides the distribution of liquidity diffusion ($\beta^{\ell}_{\sigma_{TT}}$) for each individual asset over the entire sample period. Column 1 presents the histograms, Column 2 shows the scatter plots of ($\sigma_{TT}$ vs. $\sigma^{\ell}_{TT}$) overlapping a straight line with coefficient of 1.0, and Column 3 provides the 3D scatter plots of ($\sigma_{TT}$-$\sigma^{\ell}_{TT}$-$\beta^{\ell}_{\sigma_{TT}}$) with its 2D projection ($\sigma_{TT}$-$\sigma^{\ell}_{TT}$).

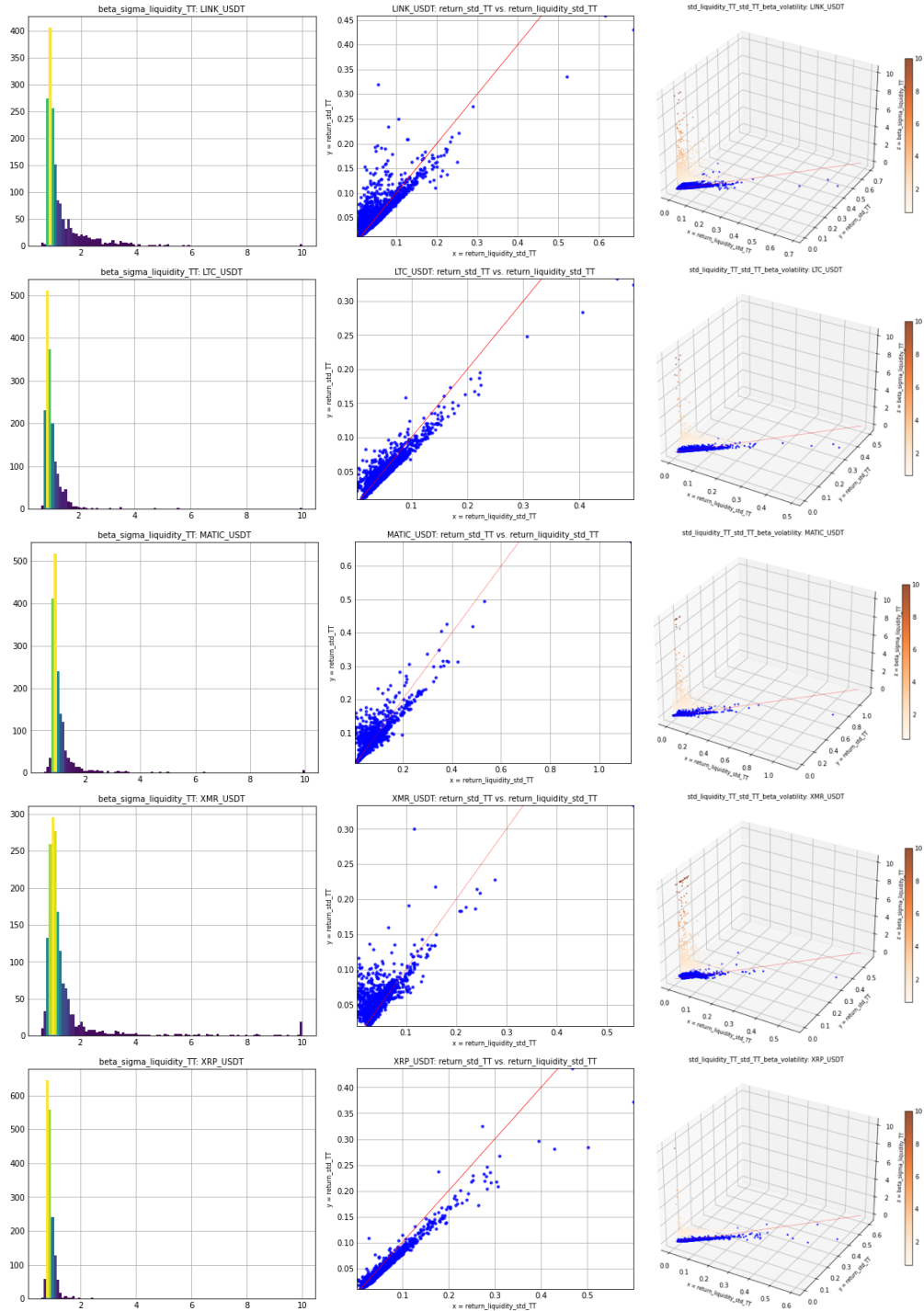